# Confinement Effects on the Crystalline Features of Poly(9,9-dioctylfluorene)

Jaime Martin[1,*], Alberto Scaccabarozzi[1], Aurora Nogales[2], Ruipeng Li[3], Detlef-M. Smilgies[3], Natalie Stingelin[1,*]

[1] Centre for Plastic Electronics and Department of Materials, Imperial College London, Exhibition Road, London, SW7 2AZ

[2] Instituto de Estructura de la Materia IEM-CSIC, C/ Serrano 121, Madrid 28006, Spain

[3] Cornell High Energy Synchrotron Source, Wilson Laboratory, Cornell University, Ithaca, New York 14853, USA

**Abstract**

Typical device architectures in polymer-based optoelectronic devices, such as field effect transistors organic light emitting diodes and photovoltaic cells include sub-100 nm semiconducting polymer thin-film active layers, whose microstructure is likely to be subject to finite-size effects. The aim of this study was to investigate effect of the two-dimensional spatial confinement on the internal structure of the semiconducting polymer poly(9,9-dioctylfluorene) (PFO). PFO melts were confined inside the cylindrical nanopores of anodic aluminium oxide (AAO) templates and crystallized via two crystallization strategies, namely, in the presence or in the absence of a surface bulk reservoir located at the template surface. We show that highly textured semiconducting nanowires with tuneable crystal orientation can be thus produced. Moreover, our results indicate that employing the appropriate crystallization conditions extended-chain crystals can be formed in confinement. The results presented here demonstrate the simple fabrication and crystal engineering of ordered arrays of PFO nanowires; a system with potential applications in devices where anisotropic optical properties are required, such as polarized electroluminescence, waveguiding, optical switching, lasing, etc.

# 1. Introduction

Understanding how the microstructure of polymers develops in spatial confinement continues to be a major fundamental issue in the field of soft condensed matter [1-5]. With the rapid rise of nanotechnologies, elements and devices based on polymeric components organized at the sub-micron scale begin to play an important role in a variety of fields, such as photovoltaics, lighting, tissue engineering, sensing, information storage, and more. [6-10]. Hence, the understanding and manipulation of the crystalline features of nanoconfined polymer materials emerge as a major issue for the nanodevice optimization. Note that the mechanical, optical, chemical and/or electrical properties of the device are directly correlated with the crystallinity, crystalline phase, size of crystals, orientation, defects, etc.

Active components of polymer-based optoelectronic devices are clear examples of functional systems where the polymer material –a semiconducting polymer- is processed into nanoscale architectures. Thus, typical device geometries in polymeric field effect transistors, organic light emitting diodes and organic photovoltaic cells include sub-100 nm semiconducting polymer thin-film active layers [11-13], whose microstructure is likely to be subject to finite-size effects. However, little attention has been hitherto paid to the consequences of the spatial limitation on the structural development in semiconducting polymer nanostructures, even though it is well accepted that the microstructure determines to a large extent the device performance [14, 15].

In the last years, the system composed of polymers confined into anodic aluminum oxide (AAO) nanopore arrays has stood out as an efficient tool to study the impact of nanoscale confinement on the crystallization of commodity polymers [16-23]. AAO nanopores constitute an ideal confining medium to assess this issue, as the degree of confinement can be easily tuned by varying the pore diameters from 10 to 400 nm [24-27]. Moreover, due to the well-defined cylindrical pores with rigid walls, the polymer melt can be considered to be effectively confined in a two-dimensional geometry. This is a significant advantage compared to templates like controlled porous glasses (CPG) having interconnected tortuous channels, or cylinder forming block copolymers, in which crystallization often dominates over microphase separation, so that the nanoscopic domain structure does not efficiently confine the crystallizing component. Furthermore, pores in AAO are arranged into a well-aligned hexagonal array, which permits investigating orientation.

Several studies have been published which include basic structural characterization of semiconducting polymer nanostructures prepared from AAO templates [28-31]. Nevertheless, most of these have focused on low aspect-ratio (length to diameter) nanostructures, whose crystalline features are known to be largely influenced by the supporting substrate. Recently, Martín et al. reported the only systematic study to date about the internal structure of a semiconducting polymer (poly(3-hexylthiophene), P3HT) crystallized from the melt inside high aspect-ratio AAO nanopores [17]. Their study revealed a strong impact of the 2D-confinement on the structure development of P3HT. For instance, they observed the stabilization of an uncommon polymorph, i.e. the form II, under strong confinement conditions, as imposed by 15 and 25 nm in diameter pores. Furthermore, they found that the crystal texture of the nanowires varied as a function of the degree of confinement (pore diameter): In large diameter pores (above 100 nm), crystals were oriented laying the π-π stacking direction parallel to the long axis of the nanopores, as corresponding to texture governed by kinetics selection rules of the crystal growth process. Conversely, inside sub-100 nm pores, the strong confinement led to nanowires in which the π-π stacking direction pointed normal to their longitudinal axis. Along the same line, it is worth noting the work by O'Carroll et al., which showed that the poly(9,9-dioctylfluorene) (PFO) melt crystallized aligning polymer chains with the pore axis when confined in commercial AAO filters (200 nm in nominal pore diameter) [32]. Such orientation is extremely uncommon for polymers confined in nanopores (it has just been observed for PFO) and its origin is not well understood yet. However, it provokes highly desired anisotropic optical properties in the nanowires [33]. Thus, these two works suggest that semiconducting polymers in confinement may behave different compared to commodity polymers, which highlights that more work is needed to elucidate how the microstructure of these materials is developed under spatial restriction.

Hence, we set out to address the effect of 2D-confinement on the internal structure of the semiconducting polymer PFO (Figure 1a), with the ultimate aim to predict and manipulate the crystalline features of this semiconducting polymer. For that, we have investigated PFO nanowires produced via two processing routes: i) First, the crystallization of confined PFO was carried out while connected by a surface bulk reservoir located at the template surface (Figure 1b). ii) Second, the confined PFO was crystallized while isolated in each nanopore (in the absence of the surface film reservoir, Figure 1c). Thus, we show that we are able to (i) fabricate highly texturized semiconducting nanowires and to (ii) manipulate their crystal orientation. Moreover, our results indicate that employing the appropriate crystallization conditions extended-chain crystals can be formed in confinement. The results presented here provide a convincing demonstration of the simple fabrication and crystal engineering of ordered arrays of PFO nanowires; a system with a clear potential for applications where anisotropic optical properties are needed, such as polarized electroluminescence, waveguiding, optical switching, lasing, and more [6, 33-36].

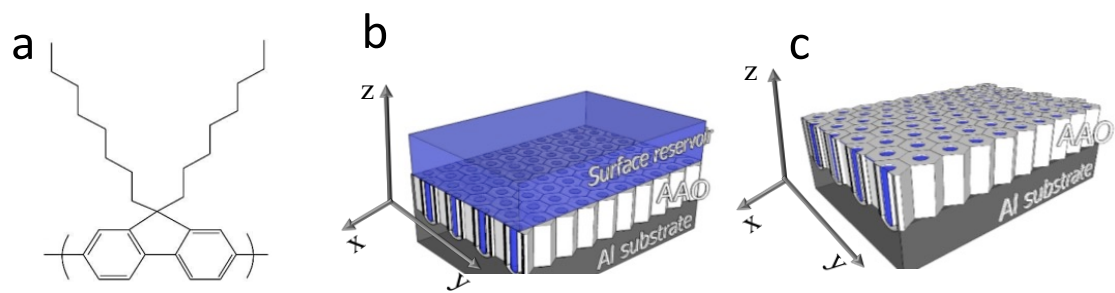

**Figure 1**. (a) Chemical structure of poly(9,9-di-n-octyl-2,7-fluorene) (PFO). Illustrations of an AAO template containing PFO nanowires connected by a macroscopic polymer reservoir (b) and an AAO template containing PFO nanowires isolated in each nanopore (c). The AAO templates have been coloured in white, while the PFO appears in blue. Axis indicating analysed directions are also included.

## 2. Experimental

*Materials*. Bulk PFO was purchased from American Dye Source, Quebec, Canada (cat. No. ADS129BE), and used as received. The chemical structure of PFO is illustrated in Figure 1a. The weight-averaged molecular weight, $\overline{M_w}$, of the polymers used in this work was estimated to be 24,000 g/mol by Chen et al. [37] based on light scattering results of Grell et al. [38]. The AAO templates were purchased from Smart Membranes GMBH, Halle, Germany, and were treated as indicated below.

*Infiltration of PFO into the Nanopores*. The AAO templates were first sonicated in different polarity solvents (hexane, acetone, isopropanol and water), in order to remove organic molecules attached to the pore walls, which would decrease the surface energy of the walls. For the infiltration, solid PFO pieces were placed on the template surface and annealed at 200 °C for 4 h under vacuum. Due to the large surface energy difference between the hydroxilated pore walls of the AAO template and the PFO melt, the polymer spontaneously wets the pore walls, which triggers the complete infiltration of the molten PFO in the nanopores [21, 39-42]. One set of samples were non-isothermally crystallized at a cooling rate of -0.5 °C/min while the PFO contained inside the pores was in contact with a macroscoscopic drop of molten PFO that remained on the template surface (on the following denoted as surface bulk reservoir). For the analysis, the residual bulk surface reservoir was removed from the template surface with a sharp razor blade and a paper tissue slightly damp with toluene. A second set of samples was rapidly quenched in ice-water immediately after the infiltration process. Subsequently, the surface bulk reservoir was removed as indicated above and, finally, the samples were annealed at 200 °C for 10 min and crystallized at 50 °C/min.

*Characterization.* The morphology of the AAO templates was characterized by scanning electron microscopy (SEM, Hitachi SU8000).

Wide-angle X-ray scattering measurements were performed at D-line, Cornell High Energy Synchrotron Source (CHESS) at Cornell University. A wide band-pass (1.47%) X-ray beam with a wavelength of 1.155Å was shone on the samples with incidence angles between 0.5° and 1°. A Pilatus 200k detector with a pixel size of 172μm was placed at a distance of 28.9 cm from the samples. A 1.5 mm wide tantalum rod was used to block the intense scattering in the small-angle area. The exposure time for room temperature measurements was 1s. The nanowire arrays were placed in the set up so that the PFO nanowires were aligned with the $z$-direction (Figures 1b and 1c). For the temperature resolved WAXS measurements, a hot stage was employed and heating and cooling rates of approx. 50 °C/min were used.

Additional X-ray diffraction θ/2θ scans were performed to at room temperature using a PANalytical X'Pert Pro MPD employing the Cu Kα radiation. The analysed wave vector $q$ was parallel to the long axis of nanowires.

Differential scanning calorimetry (DSC) was performed on a Mettler–Toledo DSC 1 Stare system at a heating/cooling rate of 20 °C/min.

Transmission optical microscopy was performed on an Olympus BX51 (withLMPlanFL N 503 objective) in combination with a temperature-controlled microscope stage (Linkam LTS 420). Micrographs were taken with a Canon 5D Mark II.

## 3. Results and discussion

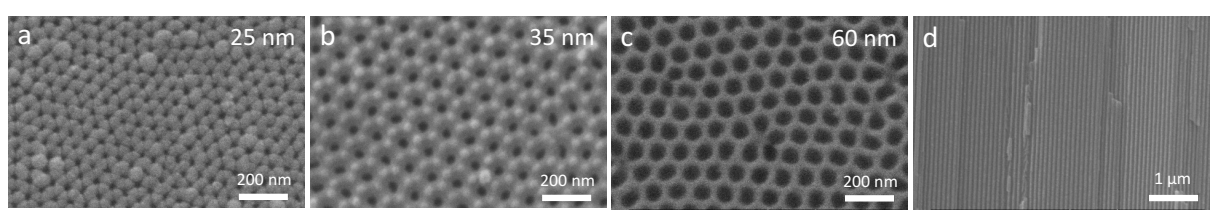

**Figure 2**. SEM images of the surface of AAO templates having 25 (a), 35 (b) and 60 (c) nm in diameter pores. (d) Cross section view of the AAO template showing the perfectly aligned array of nanopores.

The prepared samples consisted of hexagonal arrays of 25, 35, and 60 nm in diameter PFO nanowires embedded into AAO templates. The birds-eye-views and cross section of the used templates are shown in Figure 2. Thus, we explore the influence of the crystallization strategies applied on the crystalline features of the confined PFO.

## 3.1 Crystallization of the PFO nanowires in contact with the bulk surface bulk reservoir

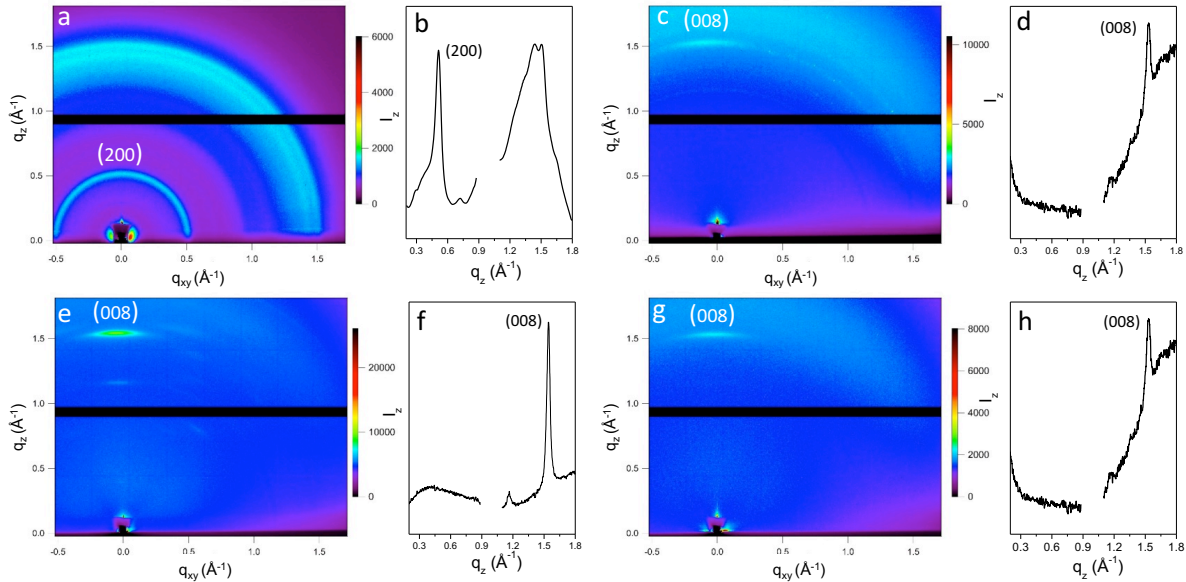

**Figure 3**. 2D-WAXS patterns of bulk PFO (a) and PFO confined inside 60 nm (c), 35 nm (e), and 25 nm (g) in diameter pores crystallized in the presence of the surface reservoir. (b, d, f, and h) correspond to the intensity collected along the *z*-direction (parallel to the long-axis of the nanopores) for the bulk, 60 nm, 35 nm, and 25 nm, respectively.

Figure 3 shows the 2D-WAXS patterns of bulk PFO (Figure 3a), and PFO confined inside 60 nm (Figure 3c), 35 nm (Figure 3e), and 25 nm (Figure 3g) pores (in diameter) crystallized with the bulk surface reservoir. All the samples, including the nanowires, show clear Bragg reflections, which indicates a semicrystalline nature of PFO, despite of being crystallized under strong confinement conditions. This characteristic is well-illustrated by the 2D-pattern of the 35 nm in diameter PFO nanowires, which shows multiple, narrow reflections in a fibre-pattern-like diffractogram. In contrast, bulk PFO shows diffraction rings, corresponding to an isotropic distribution of crystals. Intensity along the meridian (*z*-direction) reveals a diffractogram characterised by a prominent diffraction maximum at $q = 0.51$ Å$^{-1}$, large amount of diffuse scattering and few Bragg reflections. According to Chen et al. [37], such a diffractogram can be ascribed to α' phase crystals of PFO. They described the α' phase as a kinetically favoured modification of the orthorhombic α form [37]. However, its metastability [37] and high degree of disorder resemble closely a solid state mesophase rather than a real crystalline structure. Polymeric solid state mesophases are extremely frequent and are typically characterized by long range order in the parallel arrangement of chain axes (represented here by the (200) peak at $q = 0.51$ Å$^{-1}$) and a large amount of structural disorder [43]. For example, mesophases have been identified in stretched polyethylene (PE)[44], poly(ethylene terephthalate) (PET)[45], syndiotactic polystyrene [46] or in quenched isotactic and syndiotactic polypropylene[47, 48] for instance. Indeed, the PFO itself possesses a well know solid state mesophase, i.e. the β phase [38, 49-51].

Conversely, WAXS patterns of the PFO nanowires were characterized by the presence of well defined, discrete reflections that indicate high degree of orientation of PFO crystals inside the nanopores. Independently on the pore diameter, the diffractograms are characterised by a single diffraction maximum along the *z*-direction, which is parallel to the long axis of the nanopores (Figures 2d, 2f, 2g) (see also Figure 1b). These reflections can be assigned to the α phase using the indexing given by Chen et al. [52]. The peak is centred at $q=1.54$ Å$^{-1}$ (*d*-spacing of 0.416 nm) and can be identified as the stacking of (008) lattice planes with a *d*-spacing of 0.415 nm; commensurate with the interphenylene spacing along the aromatic backbone (i.e., half of the repeating unit of the polymer) [49, 52]. Such a diffractogram indicates that the polymer chains in the crystal (the *c*-axis) lay along the direction of the long axis of the nanopores.

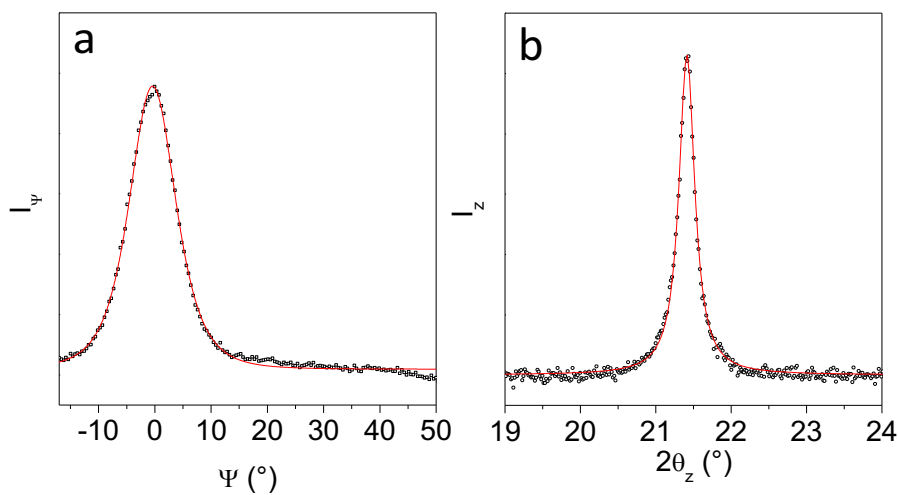

**Figure 4**. (a) WAXS intensity of the (008) peak of the ensemble of 35 nm in diameter PFO nanowires along the Ψ angle (defined as the angle between the z direction and the x-y plane). (b) Detail of the (008) peak of the 35 nm in diameter PFO nanowires as measured in θ-2θ configuration (black dotted line). Corresponding fits to Lorentzian functions are shown as red solid lines.

The orientational distribution of crystals along the Ψ angle –defined as the angle produced by the rotation of the sample normal from the *z* direction to the *x-y* plane- was evaluated from the 2D-WAXS intensity at $q=1.54$ Å$^{-1}$, i.e. the (008) reflection. Figure 4a depicts the intensity collected (black squares) for the 35 nm in diameter nanowires together with the fit of the experimental data to a Lorentzian function (red line). This sample was deliberately selected due to its high crystallinity (as deduced from the WAXS results in Figure 3e), but the presented behaviour is representative of the whole set of samples. The plot is characterized by a maximum at Ψ=0° -i.e. the *z*-direction and the pore long axis direction- and a full-with-at-the-half-maximum (FWHM) of 9.7 °, which clearly indicate that polymer chains are well aligned, being axially oriented along the nanopores. Indeed, such a low value of FWHM is usually consistent with an exclusive crystal orientation [18, 53].

It must be pointed out here that the texture observed in our nanowires is uncommon for polymers crystallized inside cylindrical domains; especially when these are crystallized under conditions promoting low density of nucleation centres, like in the present work (crystallization was induced at -0.5 °C/min). Within cylindrical nanopores, polymers tend to crystallize along one or several [*hkl*] directions with zero *l*-index, which lay parallel to the long axis of the pore [16, 18, 53-56]. The *l*-index is typically associated with the *c*-axis of the crystal, which is parallel to the polymer chain direction in the unit cell by convention. Thus, crystals typically grow along the pores with the polymer chains perpendicularly to the pore long axis.

The main reason for this behaviour is the inability of common polymer lamellar crystals to propagate along the chain direction due to the folding of chains. Consequently, the propagation is confined to the lateral directions of lamellae, i.e. the plane perpendicular to the chain direction. Hence, the alignment of the chains normal to the pore axis allows lamellae to propagate straight on the pore. Therefore, crystals having the chains oriented normal to the pore axis become larger in size and, eventually, their statistical presence becomes majority. That means that, in general, the crystal orientation within nanopores tends to be governed by kinetic selection rules of the crystal growth process. Hence, the tendency to develop crystals with chain oriented 90° to the pore axis, typically increases as long as the crystal growth process is favoured over the crystal nucleation process.

Note that these are precisely the conditions in which our nanowires are being crystallized. When the polymer in the pores is crystallized in contact to the surface reservoir at low cooling rates, crystal nuclei developed earlier in the bulk surface reservoir than in the nanopores upon cooling (due to statistical reasons). Then, nuclei form stable crystals that grow isotropically in the bulk reservoir along the crystallographic direction having the fastest growth rate, forming thus the typical spherulites. Note that such direction needs to fulfil the cero *l*-index condition. In this way, some of these crystals growing along the fastest growth direction are able to propagate through the polymer contained into the pores, in such a way that the crystallization confined melt occurs exclusively by crystal growth process, as no nucleation events have taken places inside the pores. Polymer nanowires thus crystallized frequently exhibit uniaxial crystal orientation where the crystallographic direction having the fastest growth rate – it fulfils the cero *l*-index- orients parallel to the pore axis [16, 19, 53]. For most of π-conjugated polymers, including PFO [57], the fastest growth direction is close to the direction of the π-π stacking [58, 59]. Hence, one would have expected to observe PFO crystals with the π-π stacking direction parallel to the pore axis. In contrast, our nanowires crystallized with the polymer chains parallel to the pore axis.

In order to gain information about the length-scale of such anomalous orientation, the coherence length of the (008) reflection was calculated [60]. Since the (008) lattice plane is normal to the chain direction, its coherence length addresses the crystal size along such direction and, therefore, it can be related to the thickness of the lamellae. For the analysis, the ensemble of 35 nm in diameter PFO nanowires was measured in *θ-2θ* configuration in such a way that the analysed scattering vector, *q*, was parallel to the pore axis ($q=q_z$). Note that although the experimental broadening was neglected, the resolution in *q* is better than in the GIWAXS set up used for the texture analysis. The measured FWHM accounted to 0.25° corresponding to a coherence length of 23 nm along the aromatic chain direction (Figure 4b). Since the length of the repeating unit in PFO has been calculated to be 0.83 nm [52], one can roughly estimate that the crystals possess the striking minimum value of 28 repeating units aligned in the pore direction. Such value lays between the number-average degree of polymerization ($\overline{X_n}$ = 24) and its weight-averaged counterpart $\overline{X_w}$ (= 62), which suggests that a significant fraction of the crystals consist of entire molecules. That is, PFO nanowires crystallized with bulk surface reservoir seem to be composed of extended-chain crystals. Indeed, following Wunderlich's nomenclature, one should better use the term "fully-extended-chain crystal" to refer to such structures, as the crystal dimensions in the chain direction is identical to the length of the polymer chain [61]. Unfortunately, the (008) reflection was not sufficiently well resolved for bulk PFO, which prevented the comparison between the samples. However, it is widely accepted that bulk PFO of this $\overline{X_w}$ crystallizes as folded-chain crystals [52, 62].

On account of the fact that extended chain crystals are rather uncommon entities in polymers, we tried to further characterize the lamellar thickness in our PFO nanowires with other methods, such as small angle X-ray scattering (SAXS) and DSC. However, the data collected by SAXS were not conclusive in part due to the poor contrast in this material (results not shown). Likewise, we tried to assess the thickness of lamellae by analysing the melting temperatures of crystals ($T_m$). According to the Gibbs-Thomson relation [63] (equation 1), small variations of the lamellar thickness lead to crystals with significantly different melting temperatures [64].

$$\Delta T_m = T_m^0 - T_m = \frac{2\,T_m^0}{\rho \Delta H_{f,m}} \left( \frac{\gamma_1}{L_1} + \frac{\gamma_2}{L_2} + \frac{\gamma_3}{L_3} \right) \qquad (1)$$

$T_m^0$ is the melting temperature of an infinitely large crystal, $\rho$ is the crystal density, $\Delta H_{f,m}$ is the heat of fusion per unit of mass, $\gamma_1$, $\gamma_2$, and $\gamma_3$ are the surface energies of the crystal faces, and $L_1$, $L_2$, and $L_3$ are the thickness, width, and length of the lamellae, respectively. Given that lateral dimensions of bulk polymer crystals can be considered infinite, the second and the third term of the bracket in equation (1) can be typically neglected [65]; which is why the melting point depression of bulk polymers is generally related just to the reciprocal lamellar thickness ($L_l^{-1}$) and the free energy of the fold surfaces ($\gamma_1$).

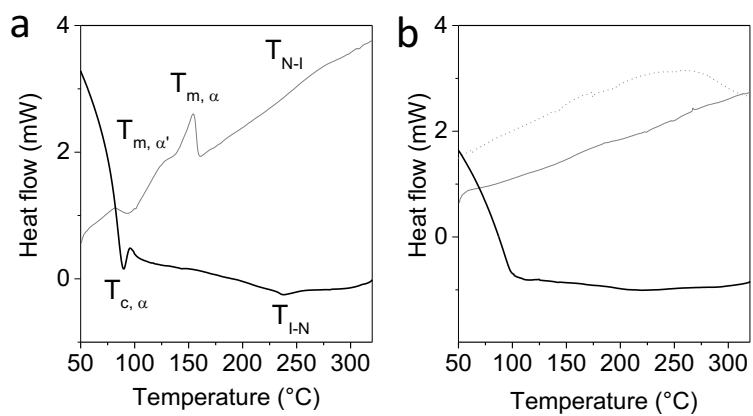

**Figure 5.** Representative heating (grey lines) and cooling (black lines) DSC traces of bulk PFO (a) and PFO inside 35 nm in diameter pores (a). Solid line and dotted line in (a) correspond to the heating run of PFO nanowires crystallized in the presence and absence of the surface reservoir, respectively.

DSC experiments were carried out on the PFO confined in 35 nm pores as well as on the bulk PFO for comparison. Results of the bulk and the nanowires are shown in Figure 5a and 5b, respectively. Although 35 nm in diameter PFO nanowires were selected due to their high crystallinity (as deduced from WAXS, Figure 3e), just a weak endotherm can be observed in the DSC heating trace (Figure 5a, dotted line), which renders any potential $T_m$ analysis difficult. Note that, uncorrected, raw DSC data are presented here in order to provide the reader with an unaltered vision of the behaviour of this system, as any correction of such a weak signal would certainly modify its shape. DSC traces of bulk PFO, in contrast, clearly shows the different thermal processes taking place in the polymer. Thereby, upon heating, one can observe an exothermic process associated with cold crystallization –at ~100 °C, and two overlapped endothermic processes, ascribed to the reorganization of α' crystals into α crystals (at $T_{m,\alpha'}$) and the subsequent melting of the α crystals into the nematic phase (at $T_{m,\alpha}$) [37, 51, 62] (Figure 5b grey line). A weak, broad peak ascribed to the nematic to isotropic transition ($T_{N-I}$) is also visible at high temperatures. The corresponding isotropic-to-nematic transition ($T_{I-N}$) is, however, better appreciated as cooling. Upon further cooling the crystallization of bulk PFO crystals takes place at ~90 °C (peak temperature).

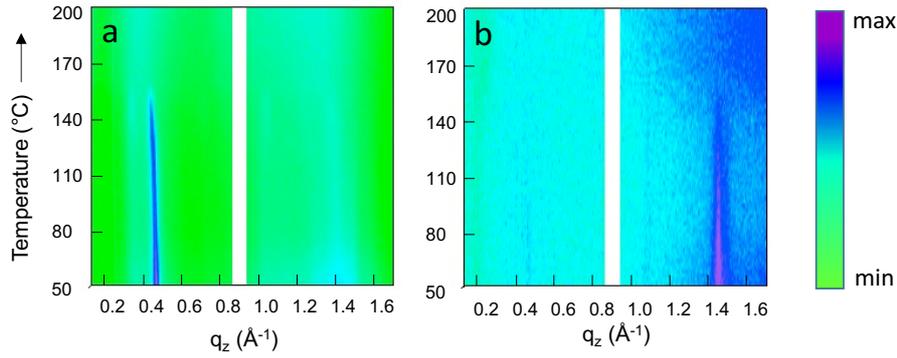

**Figure 6**. WAXS signal along the *z*-direction upon heating for bulk PFO (a) and 35 nm in diameter nanowires crystallized with surface reservoir (b). Intensity is plotted as colour scale, the temperature is shown in the vertical direction and $q_z$ in the horizontal axis).

Faced with the difficulty of analysing melting temperatures of our confined PFO crystals by DSC, temperature resolved WAXS experiments were performed. In order to compare just the impact of the dimensions of crystals on the $T_m$ and thus to avoid effects ascribed to the distinct crystallographic phases, the bulk crystals were transformed into the α phase by thermal annealing at 130 °C for 10 min, as reported by Chen et al. [37]. Figure 6 depicts the WAXS signal along the *z*-direction as a function of temperature (as heating) and $q_z$ for the bulk sample (Figure 6a) and the 35 nm in diameter nanowires crystallized with surface reservoir (Figure 6b). WAXS intensity is represented in a colour scale, whereas temperature and $q_z$ are presented along the vertical and the horizontal axis.

In contrast to the typical behaviour of nanoconfined polymers, the PFO nanowires show higher $T_m$ ($T_{m,35nm}$ = 151 °C, Figure 6b) than the bulk counterpart ($T_{m,bulk}$ =148 °C, Figure 6a). Given that, the only crystal dimension that impacts the melting temperature of bulk crystals is the one along the chain direction, the observed increase of the $T_m$ in the nanowires must be correlated to the increase of the lamellar thickness. Moreover, this thickening must be large because its effect on the $T_m$ is probably underestimated in this case. The reason is that whereas the melting point depression of bulk polymers can be just related to the reciprocal lamellar thickness ($L_l^{-1}$) and the free energy of the fold surface ($\gamma_1$), this simplification is not usually valid for polymer crystals confined in nanopores [2, 4]. Given that in our nanowires, the *c*-axis of the crystal cell (the extended chain direction) points in the direction of the nanopore axis, the other two lateral dimensions of the lamellae lay along the section of nanopores and, therefore, have nanoscale dimensions too. As a consequence, the terms containing the lateral surface energies, $\gamma_2$ and $\gamma_3$ in equation 1, are no longer negligible; contributing to the reduction of $T_m$. Indeed, notable melting point depressions are measured for polymers confined in AAO nanopores of similar dimensions [20, 54, 66, 67]. Hence, in our PFO nanowires, the increase of the lamellar thickness must be very significant, as it is able to compensate (in fact overcome) the reduction of $T_m$ induced by the pore-confinement along the other two spatial dimensions. Note that the lateral confinement of crystals prevents the employment of the Gibbs-Thomson equation (and the values of $T_m^0$, $\Delta H_{f,m}$ and $\gamma_1$ reported by Chen and coworkers) to determine the lamellar thickness and prove whether or not extended-chain crystals have been achieved, as the values of the other two interfacial energies are unknown for the PFO [62].

It is important to note in this context that, traditional methods to produce extended chain crystals from the melt imply the isothermal crystallization of the polymer under high pressures or isothermal crystallization during mechanical deformation [61]. None of these procedures have been employed here, as our PFO nanowires were crystallized non-isothermally under atmospheric pressure. Thus, the confinement in AAO pores might emerge as an additional method to produce extended-chain crystals in these kinds of polymers. This can be especially relevant for high-performing, rigid-rod semiconducting polymers, such as poly(2,5-bis(3-alkylthiophen-2-yl)thieno[3,2-b]thiophenes) (PBTTT)[68], diketopyrrolopyrrole (DPP) based polymers [69], etc. in which optoelectronic properties are directly correlated with their packing and internal microstructure.

Whether or not the mechanism for the formation of extended-chain crystals under isothermal conditions is still valid for the extended-chain crystals developed under confinement is an open question. It is well accepted that the growth of extended-chain crystals under pressure can be described as a two-step process [61]. The first step is the transition of the molecule from mobile conformations in the melt to a folded chain in a crystal [70]. Such initial crystallite may be even of metastable nature [71]. In the second step, chains reorganize into the extended-chain conformation. This reorganization takes places in the solid state; induced by the reduction of the volume provoked by the application of high pressures [72]. Hence, the absence of high pressures in our system makes such a solid state reorganization of chains unlikely. We thus can speculate that chains may get directly attached to the growth front of the extended-chain crystals already in an extended configuration in a sort of epitaxial process (due to lattice mismatch). Such a direct incorporation into an extended conformation would require either that the disentanglement time is shorter than the time needed for the translation of the chain from the melt to the crystal growth front, or directly the absence of entanglements in the liquid PFO prior to crystallization. In principle, we are inclined to the second possibility: Firstly, because it is well known that the density of entanglements is low in main-chain polymer nematics; and, secondly, because a recent work has demonstrated the dilution of the entanglement network in polymer melts under strong spatial restriction, as that imposed by the nanopores of AAO [73].

On the other hand, the question why the PFO nanowires slowly crystallized with bulk surface reservoir have a texture which is not governed by kinetic selection rules might be related to the extended nature of the crystals. Note that the lamellar thickness of these crystals is at least 23 nm, which directly conflicts with the pore diameter (35 nm). In this way, crystals with chains oriented perpendicular to the pore axis (as kinetic selection rules of the crystal growth would dictate) cannot be accommodated in the pore due to space reasons. Thus, the only orientation allowing extended-chain crystals in the pores would be the one in which the chains point in the long axis direction of the pores. The propagation of these crystals could take place through the the nucleation of new crystals on the (001) surface of the lamellae, as Wunderlich et al. have demonstrated for other extended chain crystals [70]. This would allow the propagation of the crystallization of the nanowires along the $c$-axis direction, and eventually, would explain the anomalous orientation observed in our PFO nanowires.

### 3.2 Crystallization of the PFO nanowires isolated inside nanopores

We analyse now the texture of the PFO nanowires crystallized while isolated in each nanopore, so that the crystallization occured independently in each nanopore. Thus, both nucleation and the crystal growth process have to take place under space limitation. To achieve this, samples were quenched in ice-water immediately after the infiltration process and the surface bulk reservoir was removed. The nanowires were then taken to the nematic phase at 200 °C for 10 min and crystallized at 50 °C/min.

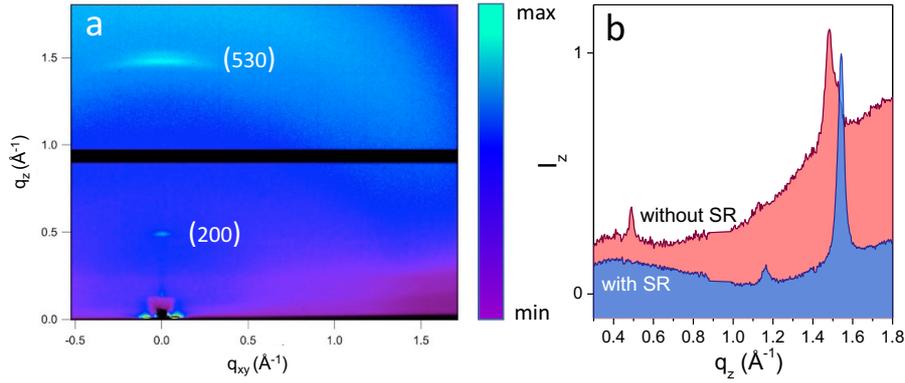

**Figure 7**. (a) 2D-WAXS patterns of the ensemble of 35 nm in diameter PFO nanowires crystallized without surface reservoir. (b) WAXS intensity along the *z*-direction (parallel to the long-axis of the nanopores) for 35 nm in diameter nanowires crystallized in the presence (blue) and in the absence (red) of surface reservoir. Intensity has been normalized to the peak value and then shifted for the clarity of the plot.

The 2D-WAXS patterns are again characterized by well-defined, main discrete reflection at moderately high *q*-values along the meridian, i.e. pore axis direction (Figure 7a). However, a closer view reveals that the texture of these nanowires is substantially different from that of nanowires crystallized with the surface reservoir (Figure 7b). The main peak observed is centred at $q=1.48$ Å$^{-1}$ (*d*-spacing of 0.425 nm) and can be ascribed to the stacking of (530) lattice planes of the α crystalline phase with a *d*-spacing of 0.427 nm. A less intense reflection, corresponding to the stacking of (200) lattice planes, can be also observed at $q=0.49$ Å$^{-1}$. Thus, the majority of the crystals point their [530] crystallographic direction parallel to the pore axis, although a small crystal fraction exists, in which the [200] direction is aligned to the pore axis. Both crystallographic directions have in common that they are normal to THE chain direction in the lattice cell. Therefore, they both correspond to the [*hkl*] direction with zero *l*-index. As mentioned before, crystal orientations in which [*hkl*] directions with a zero *l*-index lays preferentially along the pore axis direction are common within nanopores because they enable the folded-chain lamellae to grow straight in the pores.

Note that nanowires crystallized without the surface reservoir show crystal orientations that differ 90 ° from that of the same nanowires crystallized in the presence of the surface reservoir. It is striking that under ideal crystallization conditions for the growth of crystals, i.e. the presence of surface bulk reservoir and a slow cooling rate, PFO does not show a texture governed by kinetics (as the *c*-axis orients parallel to pore axis); whereas, under crystallization conditions that do not promote crystal growth, namely no surface film and high cooling rate, the observed texture is the one that one would expect for a kinetically controlled crystallization process.

Employing the argument above, the texture observed here may imply a molecular packing into chain-folded crystals, as the chain direction is contained within the plane defined by the pore section. Thus, the space available along the chain direction is limited to a value which is similar to the molecular length; thus fully extended crystals are not allowed now. Again, we can obtain information on the thickness of the lamellae and, therefore, on the folded or extended nature of the crystals, via assessing their thermal transition. For this purpose, WAXS experiments have been demonstrated to be the best choice.

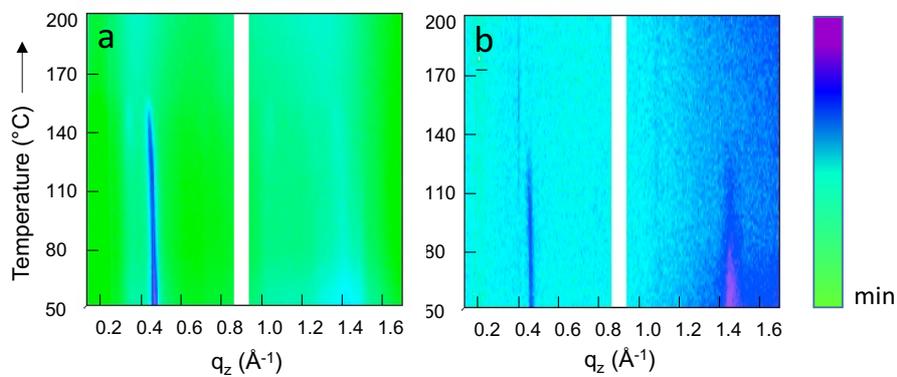

**Figure 8**. WAXS signal along the *z*-direction upon heating for bulk PFO (a) and crystallized PFO in pores (35 nm in diameter) without surface reservoir (b). Intensity is plotted as colour scale, the temperature is shown in the vertical direction and $q_z$ in the horizontal axis).

Several interesting features can be extracted from the melting behaviour of the PFO nanowires (Figure 8b) when compared to their bulk counterpart (Figure 8a). First of all, a significantly lower $T_m$ value is measured for the PFO nanowires ($T_m$ = 129 °C) than for bulk PFO ($T_m$ = 148 °C). In the nanowires crystallized without surface film, the *c*-axis of the lattice cell (the chain direction) orients perpendicular to the nanopore long axis, so that one of the lateral dimensions of the lamellae positioned normal to the pore axis and will be thus constrained, while the other lateral dimension points along the pore axis. Hence, at least one of the terms containing the lateral surface free energy contribution ($\gamma_2$ in Eq. 1) is not negligible and contributes to the melting point depression. Apart from this, the formation of thinner lamellae inside pores cannot be ruled out, which would decrease further $T_m$ [20, 67].

Let us compare now the melting temperature of the nanowires crystallized without surface bulk reservoir ($T_m$ = 129 °C) with that of the nanowires crystallized in the presence of the surface reservoir ($T_m$ = 151 °C). We suggest that such difference is basically correlated with the lamellar thickness. As stated in the previous section, nanowires crystallized with surface reservoir are comprised of lamellae that are confined in the three spatial dimensions: the thickness of lamellae ($L_1$) has always nanoscale size; but, in addition, both lateral dimensions of lamellae ($L_2$ and $L_3$) in these nanowires are also nanoscopic, as these lie in the plane of the pore section. Conversely, in nanowires crystallized without surface reservoir, one of the lateral directions ($L_2$) is contained in such plane, while ($L_3$) points along the pore. However, due to the high cooling rate applied for the crystallization of nanowires without surface reservoir (50 °C/min), the density of active nucleation centres that initiate crystallization must be very high. Thus, the length of the crystal along $L_3$ is likely to be restricted by the growth-limited crystallization. Hence, in both kinds of nanowires (i.e. crystallized with and without surface reservoir), crystals are expected to be spatially confined along the three dimensions, and thus the predominant difference in $T_m$ is likely to be influenced by the change of lamellar thickness. This assumption is further supported by the fact that the energy of the fold surface ($\gamma_1$), is typically larger than the lateral surface energies ($\gamma_1$ and $\gamma_2$) in polymer crystals.

On the other hand, a broad melting process is observed in the nanoconfined PFO, which is characterized by a progressive loss of the long-range order of the (530) planes correlation. This broad melting process can be probably associated with a large heterogeneity of the nanoconfined crystals and with melt-reorganization processes [54, 55]. However, the disappearance of the (200) occurs rather abruptly.

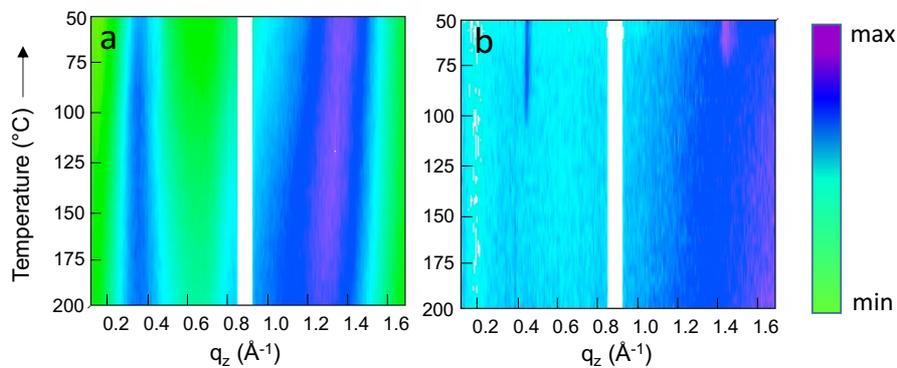

**Figure 9.** WAXS signal along the *z*-direction upon cooling from 200 °C for bulk PFO (a) and 35 nm in diameter nanowires isolated in the pores (no surface reservoir was present on the template) (b)

Likewise, a similar complex solidification behaviour can be observed upon cooling the nanoconfined PFO from the nematic state (Figure 9b). Indeed, the cooling WAXS colour plot in Figure 9b looks like a mirror image of the one depicting the heating run (Figure 8b). Upon cooling, a diffraction maximum appears in the low $q$ region at T~104 °C, which can be related to the intermolecular periodicity along the direction normal to the chain. Upon further cooling, at T~66 °C, the peak ascribed to the stacking of (530) lattice planes becomes visible.

A commonly employed argument to justify a two-step crystallization in pores assumes that both heterogeneous and homogeneous nucleation mechanisms are active, and depending on the presence or absence of heterogeneities in each individual pore, nuclei are formed via a heterogeneous or a homogeneous pathway, respectively [66, 74]. Both mechanisms are activated at different temperatures because the energy barrier that must overcome for the development of stable nuclei is different in both mechanisms. Moreover, secondary nucleation process can be responsible of the low temperature reflection.

Nevertheless, the fact that the first noticeable peak is the one corresponding to the interchain correlation –represented by the (200) reflection- leads us to consider also another possibility: the development of a non-crystalline, yet well-ordered structure at T~104 °C. This structure would be characterized by a well-defined periodicity along the $a$-axis (by analogy with the crystal cell) but absence of long-range order in other directions. Then, upon further cooling, such structure would evolve to well-defined α-crystals that would be fully visible at T~66 °C. The nature of such intermediate structure is a challenging question. The narrow (200) reflection and its sharp appearance do not look like the typical features of a polymer liquid crystalline (nematic) phase. Furthermore, the nematic phase of PFO is well known for its poorly-defined reflections in WAXS [51], as can be also deduced from Figure 8a and 9a.

However, it is well known that the periodic molecular arrangement of nematics is often altered (even enhanced) when they are confined into nanoscale pores because the orientation induced by surface anchoring and the wall-induced density modulations have lengthscales that compete with those set by elasticity and bulk correlations [75]. Furthermore, the peak could be also ascribed to a solid-state mesophase that evolves into the α-crystal form upon further cooling. This is reminiscent of the so called α' phase mentioned before, which, indeed, seems to be the structure formed in the bulk under the same cooling conditions (Figure 9a). In any case, this intermediate structure would act as a bridge between the liquid nematic phase and the α crystal and thus it would also transfer the molecular orientation from the nematic to the crystal form. In this way, the rigid PFO chains are probably already oriented perpendicular to pore axis in the nematic mesophase, and the molecular orientation is preserved during crystallization, so that, eventually, crystallographic directions perpendicular to the chain axis, i.e. [530] and [200], becomes positioned parallel to the pore axis. Note that, this hypothesis implies that the orientation observed in PFO nanowires crystallized without surface reservoir does not have a kinetic origin either, even though only [*hk0*] directions are visible along the pore axis, which would resolve the paradox of why nanowires crystallized slowly with surface film did not show a texture governed by kinetics, while fast crystallization without surface film seemed to induce kinetically controlled orientations.

**Conclusions**

We have demonstrated that the development of the internal structucture of the rigid-rod, semiconducting polymer PFO can be effectively controlled by the 2D spatial confinement and the employment of crystallization strategies. The effect of nanoconfinement on the overall crystallization process depends strongly on the individual relationship between the crystal nucleation and the crystal growth with the available space. Hence, the application of crystallization strategies that impact the nucleation and growth processes allows for the manipulation of the crystal texture of conjugated polymers in confinement and, thus, to produce semiconducting nanowires with tailored microstructures.

When PFO is crystallized inside the nanopores in the presence of a bulk surface reservoir highly crystalline and textured nanowires are achieved, in which crystals are uniaxially oriented along the chain direction in the crystal, i.e. the *c*-axis, parallel to the long axis of the nanopores. Interestingly, the coherence length of the (00*l*) reflection along the pore direction and the increase of the melting temperature of these nanowires as compared to bulk PFO crystals suggest that PFO crystallizes as extended-chain crystals inside the nanopores. The mechanism of formation and the unusual orientation of these extended-chain crystals are still open questions. However, a crystallization process where polymer chains incorporate (in a sort of epitaxial process) to the front of already extended-chain crystals might explain the formation of such an exotic structure. Then, the nucleation of new crystals would take place on the (001) surface of the lamellae, allowing the propagation of crystals along the *c*-axis direction. Note that a scenario where chains are slightly (if any) entangled in the melt (due to the strong confinement and the stiffness of the backbone chain) and the slow cooling conditions, is suitable for the development of thermodynamically preferred structures.

Conversely, PFO nanowires crystallized isolated inside the nanopores (without the surface reservoir) exhibit a preferential texture in which chains are oriented normal to the long axis of the pores. That is, the crystal orientation in these nanowires differs 90° from that of the same nanowires crystallized in the presence of the surface reservoir. These crystals are probably chain-folded, as suggested by the strong melting point depression measured. We propose that the orientation of the liquid crystal and the fast cooling rate are responsible of the crystal texture in these nanowires. Nonetheless, it is still unclear why the influence of the presence of the nematic phase on the crystal orientation is limited to the nanowires crystallized in an isolated fashion in the pores, while it is irrelevant when these are crystallized in the presence of the bulk reservoir.

The findings presented here provide, however, an excellent example of the simple fabrication and structural manipulation of ordered arrays of well-aligned semiconducting nanowires that can be achieved via confined crystallization, which may help establishing the foundation of a new range of architectures/devices with highly-anisotropic-response, such as polarized electroluminescence devices, waveguides, optical switching devices, lasers, and more.

**Acknowledgements**

Jaime Martín acknowledges support from the European Union's Horizon 2020 research and innovation programme under the Marie Skłodowska-Curie grant, agreement No 654682.The work has been partially supported under a KAUST Global Collaborative Research Academic Excellence Alliance (AEA) grant. This work is based upon research conducted at the Cornell High Energy Synchrotron Source (CHESS) which is supported by the National Science Foundation and the National Institutes of Health/National Institute of General Medical Sciences under NSF award DMR-1332208.